\documentstyle[sprocl]{article}
\begin{document}

\title{Introduction to {\em The Supersymmetric World: The 
Beginnings of The Theory}}
 \author{G. L. KANE}
 \address{Randall Laboratory, University of Michigan, \\ Ann Arbor, 
MI 48109}
 \author{M. SHIFMAN}
 \address{Theoretical Physics Institute,\\
 University of Minnesota, Minneapolis, MN 55455}

\maketitle
\abstracts{
        This is the foreword to the book we edited on the
origins and early development of supersymmetry, which has been 
just issued by World Scientific. This book presents a view on
the discovery of supersymmetry and pioneering investigations 
before summer 1976, mainly in the words of people who 
participated. It combines anecdotal descriptions and
personal reminiscences with more technical summaries of the 
trailblazers, covering the birth of the theory and its first 
years---origin of the idea, four-dimensional field theory realization,
and  supergravity. The eyewitnesses convey to us the drama of  one of the 
deepest discoveries in theoretical physics in the 20$^{\rm th}$
century. Contributors: V.~Akulov, R.~Di~Stefano, P.~Fayet, S.~Ferrara,  
G.-L.~Gervais, N.~Koretz-Golfand, E.~Likhtman, M.~Marinov, 
A.~Neveu, L.~O'Raifeartaigh, P.~Ramond, B.~Sakita, J.~Schwarz, 
M.~Sohnius, V.~Soroka, J.~Strathdee,
 D.~Volkov, J.~Wess,  P.~West.}

The history of supersymmetry is exceptional.  All other major 
conceptual developments in physics and science have occurred 
because scientists were trying to understand or study some 
established aspect of  nature, or to solve some puzzle arising from 
data.  The discovery of supersymmetry in the early 1970's, an 
invariance of  the theory under interchange of fermions and bosons, 
was  a purely intellectual achievement, driven by the logic of 
theoretical development rather than by the pressure of  existing 
data. Thirty years elapsed from the time of discovery,
immense theoretical effort was invested in this field, over 30,000 
papers published. However, none of them can claim to report the 
experimental discovery of supersymmetry (although there are some 
hints, of which  we will say more later). In this respect the 
phenomenon is  rather unprecedented in the  history of physics. 
Einstein's general  relativity, the closest possible analogy
one can give, was experimentally confirmed within several years 
after its creation. Only in one or two occasions have theoretical 
predictions of  a comparable magnitude had to wait for experimental 
confirmation that long. For example, the   neutrino had a time lag of 
27 years. 

It would not be an exaggeration to say that today supersymmetry 
dominates theoretical high energy physics. Many believe that it will 
play the  same revolutionary role in the physics of the 21$^{\rm st}$ 
century as special  and general  relativity  did in the physics of the 
20$^{\rm th}$ century.  This belief is based on aesthetical appeal, on 
indirect evidence, and on the  fact that no theoretical alternative is in 
sight.

The discovery of supersymmetry presents a dramatic story
dating back to the late 1960's and early '70's.  For young people who 
entered high energy physics in the 1990's this  is ancient history.  
Memories fade away as live participants of these events approach 
the retirement age; some of them  have already retired and some, 
unfortunately, left this world. Collecting live testimonies of the 
pioneers, and preserving them for the future, seems timely given the 
impact supersymmetry has already produced on the  development
of particle  physics. Having said that, we note that this book did not 
appear as  a result of a conscious project.  Both editors had
collected some materials for other activities\,\cite{in_2,in_3} 
and became aware of the other's interest and materials. Many people 
have been interested in how supersymmetry originated ---the 
question often is asked in informal conversations---and how it can 
be such an active field even before direct experimental confirmation.  
We finally decided to combine materials, invite further ones, and edit 
this volume that makes available a significant amount of information 
about the origins of this intellectually exciting area. Most of it is in 
the words of the original participants.

In the historical explorations of scientific discoveries (especially, 
theoretical) it is always very difficult to draw a ``red line" marking 
the true beginning, which would separate ``before" and ``after." 
Almost always there  exists a chain of works which interpolates, 
more or less continuously, between the distant past and the present. 
Supersymmetry is no exception, the  more so because it has multiple 
roots. It was observed as a world-sheet  two-dimensional 
symmetry\,\footnote{The realization that the very same string 
theories gave rise to supersymmetry in the target space came much 
later.} in string theories around 1970; at approximately the same 
time Golfand and Likhtman found the superextension of the
Poincar\'e algebra and constructed the first four-dimensional
field theory with supersymmetry, (massive) quantum 
electrodynamics of spinors and scalars. Within a year  Volkov and 
collaborators (independently) suggested nonlinear realizations of 
supersymmetry and started the foundations of  supergravity. Using 
the terminology of the string practitioners one can say 
that the first supersymmetry revolution occurred in 1970-71 as the 
idea  originated.\footnote{In the Marxist terminology it would be 
more exact to say that this was a prerevolutionary situation. This 
nuance is too subtle, however, and cannot be adequately
discussed in this article.}  The second supersymmetry revolution 
came with the work of Wess and Zumino in 1973. Their discovery 
of linearly realized supersymmetry in four dimensions
opened to the rest of the community the gates to the
Superworld. The work on supersymmetry was tightly woven in the
fabrique of  contemporary theoretical physics. During the first few 
years of its development, there was essentially no interest in 
whether or how supersymmetry might be relevant to understanding 
nature and the traditional goals of physics.  It was  ``a solution in 
search of a problem." Starting in the early 1980's, people began
to realize that supersymmetry might indeed solve some basic 
problems of our world.   This  time may be characterized as  the
third supersymmetry revolution.

So, how far in the past should one go and where  should one stop
in the book devoted to the beginnings?

The above questions hardly have unambiguous answers.
We decided to start from Ramond, Neveu, Schwarz, Gervais, and
Sakita whose memoirs are collected in the chapter entitled {\em The
Predecessors}, which opens the book. The work of these authors can 
be  viewed as precursive to the discovery of supersymmetry in 
four dimensions. It paved the way to Wess and Zumino.

Chapter 2 presenting  {\em The Discovery} is 
central in the first part of the book. It contains recollections of
Likhtman, Volkov, Akulov, Koretz-Golfand (Yuri Golfand's widow)
 and the 1999 Distinguished Technion Lecture of Prof. J. Wess, in 
which  the basic stages of the theoretical construction are 
outlined.   Chapter 3 is devoted to the 
advent of supergravity. The fourth chapter is entitled {\em The 
Pioneers}. The definition of pioneers (i.e. those who made crucial  
contributions at the earliest stage) is  quite ambiguous,  as is the 
upper cut off in time which we    set, {\em the  summer of 1976}. By 
that time  no more than a few dozen of original papers on
supersymmetry had been published. 

The selection of the contributors was a  difficult task. For various
reasons we were  unable to give  floor to some theorists who were
instrumental at the  early stages (e.g. R. Arnowitt, L. Brink, R.
Delbourgo, P.G.O. Freund,  D.R.T. Jones, J.T.~$\mbox{\L}$opusza\'nski, P.
Nath, Y. Ne'eman,   V.I. Ogievetsky, A. Salam, E. Sokat\-chev,  B. de
Wit). Some  are represented in other chapters (e.g. S. Ferrara whose 1994
Dirac  Lecture is being published    in Chap. 3.) Others are beyond
reach.  This refers to Abdus Salam and Victor Ogievetsky.
The latter, by the way, wrote (together with L. Mezincescu)  the first
comprehensive review on supersymmetry which was published in
1975.\cite{in_ogi} Even now it remains an excellent introduction to 
the subject, in spite of the 25 years that  have elapsed.

 The question of where to draw the line tortured us, and we bring 
our apologies to all the pioneers who ``fell through the cracks." 

The second part of the book is  an attempt to present a historical 
perspective on the development of the subject.  This  task obviously 
belongs to  the professional historians of science; the most 
far-sighted of them will  undoubtedly turn their attention to
supersymmetry  soon. For the time being, however,  to the best of 
our knowledge, there are no professional investigations on the issue.
There was available a treatise written by Rosanne Di Stefano in 1988 
for a conference proceedings which were never published. This is a 
very thorough and insightful  review. On the factual side it goes far 
beyond any other material on the history of supersymmetry one can 
find in the literature. There are some omissions, mostly  regarding 
the Soviet contributors, which are naturally explained by the 
isolation of the Soviet community before the demise of the USSR
and relative inaccessibility of several key papers written in Russian.
The Yuri Golfand Memorial Volume\,\cite{in_3} which contains the 
English translation of an important paper by  Golfand and
Likhtman\,\cite{in_gl} as well as a wealth of other relevant 
materials,  fills the gap. In addition, Springer-Verlag
has recently published  Memorial Volumes in honor of Dmitry 
Volkov\,\cite{in_dv} and Victor Ogievetsky,\cite{in_ogim} which 
acquaint the interested reader with their  roles to a much fuller 
extent than previously. 
 
The coverage of certain physics  issues in Di Stefano's essay required 
comment; in a few cases we added explanatory footnotes. Di Stefano's 
essay is preceded by a relatively short article written by the late 
Prof. Marinov. It is entitled ``Revealing the Path to the Superworld" 
and was originally  intended for the Golfand Volume. This article 
presents ``a bird's eye view" on the area. On the factual side it is 
much less comprehensive than Di Stefano's, but it carries a 
distinctive flavor of the testimony of an eye witness. Moreover, it 
reveals the mathematical roots of the discovery, an issue which is 
only marginally touched in Di Stefano's essay.

We are certainly not professional  historians of science; still we 
undertook a little investigation of our own. Often students ask where 
the name ``supersymmetry" came from? It seems that it was coined 
in the paper by Salam and Strathdee\,\cite{in_SS} where these 
authors constructed supersymmetric Yang-Mills theory.
This paper was received by the editorial office on June 6, 1974,
exactly eight months after that of Wess and Zumino. 
Super-symmetry (with a hyphen) is in the title, while in the body of 
the paper Salam and Strathdee use both, the old version of Wess and 
Zumino, ``super-gauge symmetry," and the new one.  An earlier 
paper of Ferrara and Zumino\,\cite{in_FZ} (received by the editorial 
office\footnote{The  editorial note says it was received  on May 27,
197{\bf 3}. This is certainly a misprint; otherwise, the event would 
be  acausal.} on May 27, 1974) where the same problem of  
super-Yang-Mills  was addressed,  mentions
only supergauge invariance and supergauge transformations. 

\newpage

Supersymmetry is nearly thirty years old. It seems that now  we are 
approaching  the  fourth supersymmetry revolution which will 
demonstrate its relevance to  nature.  Although not numerous, we do 
have hints that this is the case. They are:  (a) supersymmetry allows 
a stable hierarchy between the weak scale and the  shorter distance 
scales such as the Planck scale or unification scale, (b) 
supersymmetry provides a way to understand how the electroweak 
SU(2)$\times$U(1) symmetry is  broken, so long as the top quark 
came out heavy (which it did), (c)  gauge couplings unify rather 
accurately when superpartners are included in  the
loops,\footnote{An alternative way to say this is to say that the value 
of  the weak mixing angle at the weak scale can be calculated 
accurately if one  sets it to the value predicted by a unified theory at 
the unification scale.} (d)  the  Higgs boson is predicted to be light  
(LEP gives $M_H <200$ GeV), and (e) the lack of any deviations from 
Standard Model predictions in the precision data at LEP and 
in other experiments is consistent with supersymmetry (it was 
anticipated that these deviations  would be invisible).

Certainly,  at the moment the indications are not conclusive.
However inconclusive, they are the source of hope
and enthusiasm for  phenomenologically oriented theorists and 
experimentalists who would like to keep high-energy physics in the 
realm of empirical science.

Another aspect which came to limelight recently is the fact that 
supersymmetry became instrumental in the solution of highly 
nontrivial dynamical issues in strongly coupled non-supersymmetric 
theories, which defied solutions for decades.  That of course does not 
imply that nature is supersymmetric, but it does add to the interest 
in supersymmetry.

Summarizing, in this book we bring together contributions from 
many of the key players of the  early days of supersymmetry.
We leave its relevance to our world to a future project.

\vspace{0.5cm}
\section*{Related Materials}

Proceedings of the Symposium {\em Thirty Years of 
Supersymmetry}, Minneapolis, October 13-15, 2000, Ed. by K. Olive, 
S. Rudaz, and M. Shifman
(North Holland, 2001). 

{\em Thirty Years of 
Supersymmetry}, in Theoretical Physics Institute Newsletter \# 2,
http://www.tpi.umn.edu/Issue2.pdf.

E. Likhtman, {\em Around SuSy 1970}, Talk at the Symposium 
{\em Thirty Years of Supersymmetry,} Minneapolis,  October 13-15, 2000, 
hep-ph/0101209.
 
\end{document}